\documentclass[preprint2]{iopart}  

\usepackage{graphicx}
\usepackage{url,graphics,epsfig}
\usepackage{amssymb}
\begin{document}
\jl{2}
%
%
\def\etal{{\it et al~}}
\def\newblock{\hskip .11em plus .33em minus .07em}
%
%
%
%
%
\setlength{\arraycolsep}{2.5pt}             

\title{Rovibrationally-Resolved Photodissociaton of SH$^+$}

\author{E. C. McMillan$^{1}$, G. Shen$^{1,2}$, J. F. McCann$^{3}$
	  B. M. McLaughlin$^{3,4}\footnote[1]{Corresponding author, E-mail: bmclaughlin899@btinternet.com}$,
             P. C. Stancil$^{1}\footnote[2]{Corresponding author, E-mail: stancil@physast.uga.edu}$}
\address{$^{1}$Department of Physics and Astronomy and  the Center for Simulational Physics,
			University of Georgia, Athens, GA 30602-2451, USA}
\address{$^{2}$Institute for Applied Physics and Computational Mathematics,
			Beijing, 100094, China}
\address{$^3$Centre for Theoretical Atomic and Molecular Physics (CTAMOP),
	School of Mathematics and Physics, Queen's University Belfast,
	Belfast BT7 1NN, Northern Ireland, UK}
\address{$^4$Institute for Theoretical Atomic, Molecular, and Optical Physics (ITAMP),
	Harvard-Smithsonian Center for Astrophysics,
	 MS-14, Cambridge, MA 02138, USA}

%
%

\begin{abstract}
Photodissociation cross sections for the SH$^+$ radical are computed from all rovibrational (RV)
levels of the ground electronic state X$~^3\Sigma^-$ for wavelengths from threshold to 500~\AA.
The five electronic transitions, $2~ ^3\Sigma^- \leftarrow$ X$~^3\Sigma^-$, $3~ ^3\Sigma^- \leftarrow$ X$~^3\Sigma^-$,
$A~ ^3\Pi \leftarrow$ X$~^3\Sigma^-$, $2~ ^3\Pi \leftarrow$ X$~^3\Sigma^-$, and $3~ ^3\Pi \leftarrow$ X$~^3\Sigma^-$,
are treated with a fully quantum-mechanical two-state model, {i.e. no non-adiabatic coupling between excited states 
was included in our work.}. The photodissociation calculations incorporate adiabatic
potentials and transition dipole moment functions computed in the multireference configuration interaction approach
along with the Davidson correction (MRCI+Q), but adjusted to match available experimental molecular data and asymptotic
atomic limits. Local thermodynamic equilibrium (LTE) photodissociation cross sections were computed which
assume a Boltzmann distribution of RV levels in the X$~^3\Sigma^-$ molecular state of the SH$^+$ cation. 
The LTE cross sections are presented for temperatures in the range 1000-10,000~K. 
Applications of the current photodissociation cross sections to interstellar gas,
photon-dominated regions, and stellar atmospheres are briefly discussed.
\end{abstract}

\vspace{1.0cm}
\begin{flushleft}
Short title: Photodissociation of the SH$^+$ cation\\
J. Phys. B: At. Mol. Opt. Phys. : \today
\end{flushleft}

\maketitle

%
%
%

\section{Introduction}
In the search for novel molecular species in interstellar and circumstellar environments, 
many sulfur-containing molecules have been observed and identified in diffuse, translucent, and dense clouds
\cite{mil90,woo15}.  The 
sulfur-bearing molecules SH and SH\textsuperscript{+} have recently been observed in the interstellar 
medium (ISM) \cite{yam00,ben10}. In particular, the SH$^+$ 526 GHz transition was detected 
in emission using the {\it Herschel} infrared space observatory toward the high-mass protostar 
W3 IRS 5 \cite{ben10} and in the diffuse ISM of distant star-forming regions \cite{god12}.  
Towards Sgr B2(M), the 638 GHz transition of SH\textsuperscript{+} has also been detected 
in absorption using the Carbon Heterodyne Array of the MPIfR (CHAMP+) receiver of the 
Atacama Pathfinder Experiment 12 m telescope \cite{men11}. SH and SH\textsuperscript{+} 
facilitate the formation of other sulfur-bearing molecules, so information 
about their abundances is relevant for the investigation of the chemical composition of 
astrophysical environments.  As searches for new molecular species continue, molecular formation and 
destruction processes involving sulfur are of considerable interest.  These molecular reactions, along
with other processes, can be used to create a chemical network that can predict the 
concentrations of major sulfur species in various astrophysical environments \cite{mil90,woo15,mil86,pin86}.  
One such process is photodissociation,
an important mechanism for the destruction of molecules in the ISM.  For example, the sulfur hydride ion may be dissociated by
\begin{equation}
{\rm SH}^ + + h\nu \rightarrow {\rm S}+ {\rm H}^+.
\end{equation}
The surviving SH\textsuperscript{+} enables the formation of other sulfur-bearing molecules, 
including SH and H$_n$S\textsuperscript{+}.  In the ISM, SH\textsuperscript{+} forms 
through an endothermic reaction of H$_2$ and S\textsuperscript{+},
\begin{equation}
{\rm H}_2 + {\rm S}^ + \rightarrow {\rm SH}^ + + {\rm H},
\end{equation}
and through radiative association
\begin{equation}
{\rm S}^ + + {\rm H} \rightarrow {\rm SH}^ + + h\nu,
\end{equation}
\cite{Stancil2000}.
For the endothermic reaction, state-to-state formation rates are not available, so the UMIST 
estimations based on CH\textsuperscript{+} are often adopted.  This assumption is reasonable 
because SH\textsuperscript{+} forms via a similar reaction for CH\textsuperscript{+} and the 
total rates of the reactions for CH\textsuperscript{+} and SH\textsuperscript{+} formation 
through these processes are expected to be of the same order of magnitude.  
This endothermic reaction requires the passage of a shock, specifically a 
magnetohydrodynamic shock, due to an activiation barrier of 9960 K.  
The endothermicity of this reaction likely contributes to the difficulty of detection 
of SH$^+$ in the ISM.  In many environments, such as photon-dominated regions (PDRs), 
SH\textsuperscript{+} does not react with H$_2$ at the given physical conditions resulting 
in a reaction rate orders of magnitude lower than that of reactions with electrons and H.  
Consequently, photodissociation and the reactions with H and 
electrons dominant SH\textsuperscript{+} destruction \cite{nag13}.

 Using ab initio molecular 
 potential energies and transition dipole moment functions in a quantum-mechanical 
 scattering formalism, we have carried out the first calculations for the photodissociation of SH$^+$. These computed 
 photodissociation cross sections can be used to obtain local photo-rates which 
 can then be folded into chemical network models to improve the sulfur chemistry. In particular, the
 cross sections have been computed for five transitions to the lowest-lying electronic 
 states from 841 rovibrational (RV) levels of the X$~^{3}\Sigma^{-}$ electronic ground state.  
 Cross sections for thermal populations of RV levels are also obtained and 
 astrophysical applications of the current results are briefly discussed.  
 
%
%
%
%
%
\section{Theory}\label{secTheory}

\subsection{Electronic Structure}

Following early work on this molecular cation \cite{Stancil2000} for radiative association,
and our most recent study on radiative charge transfer \cite{Shen2015},
 we use our accurate results (obtained from a parallel version of the 
 MOLPRO~\cite{Werner2010} suite of {\it ab initio} quantum chemistry codes,
 release MOLPRO 2010) for potential energies and transition dipole moments (TDMs), 
 as well as additional TDMs, to study photodissociation of this cation from all  RV levels.
The potential energies curves and transition dipole moments were obtained as a function of bond length out to  
a bond separation of  $R=20$ a$_0$.  For internuclear separations beyond this, 
we use a multipole expansion to represent the long-range part of the potentials. 

We have previously outlined in detail the molecular structure calculations \cite{Shen2015} so here we 
just give a brief outline.  We conducted multi-reference configuration interaction using the internally contracted MRCI 
method available in the MOLPRO quantum chemistry suite of codes.  The orbitals used in this procedure were 
obtained from the State-Averaged Multi-Configuration-Self-Consistent-Field 
(SA-MCSCF) method \cite{Helgaker2000}. 
The Davidson  correction (referred to as Q) was applied to all our results \cite{Davidson1974}.
In our work the core orbitals are optimised and the valence orbitals are correlated at the CASSCF level.

For the SH$^+$  cation, the natural Abelian symmetry group is C$_{\infty v}$ but due to the limitation of the Gaussian orbitals 
code MOLPRO, our electronic structure computations were performed in the reduced Abelian symmetry group C$_{2v}$ 
i.e. (A$_1$, B$_1$, B$_2$, A$_2$). 
	%
	%
\begin{figure*}
\includegraphics[scale=0.50]{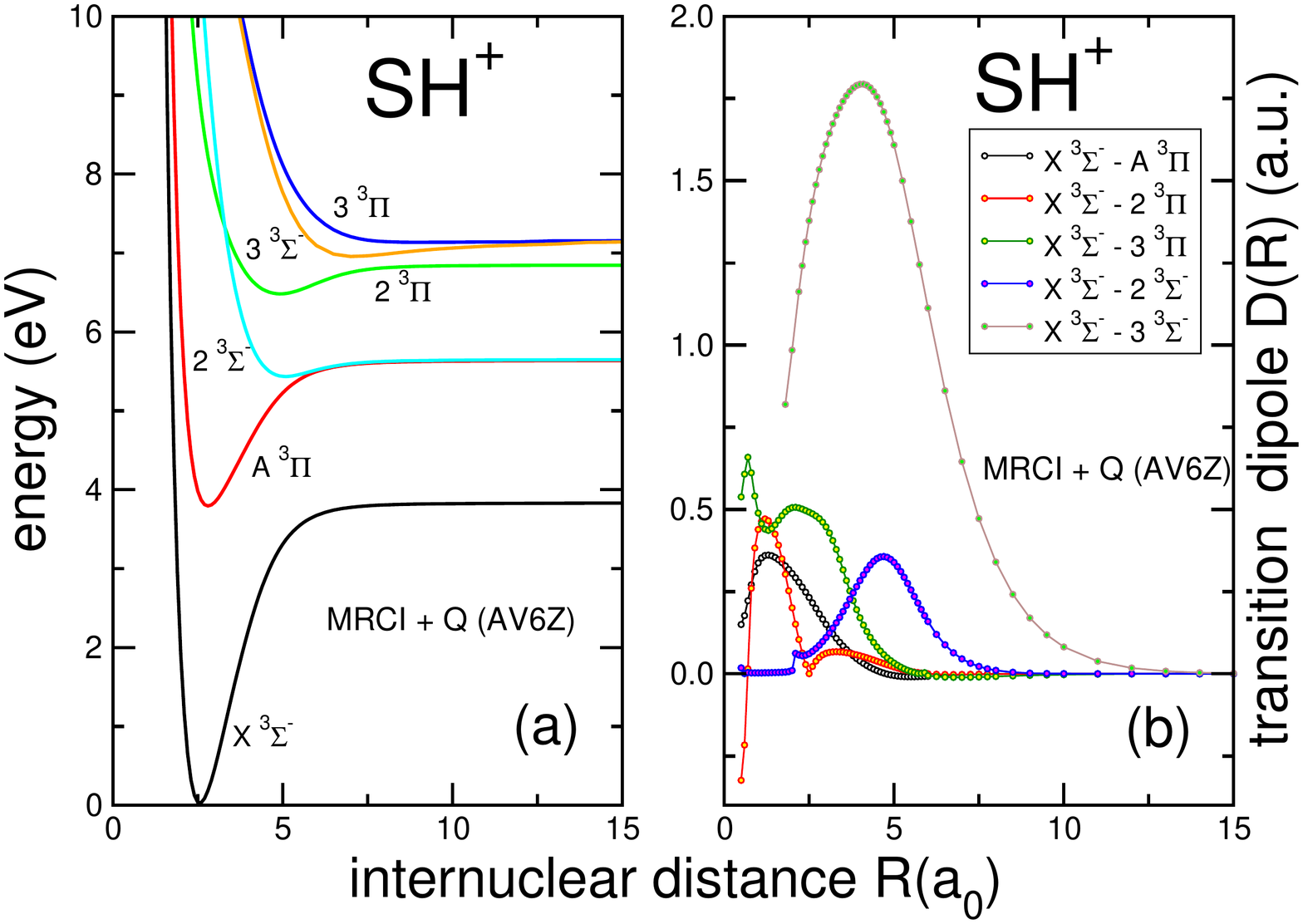}
\caption{(a) Relative electronic energies (eV) for the ${\rm SH}^+$ molecular cation, as a 
               function of bond separation at the MRCI+Q level of approximation with an AV6Z basis. 
               Energies are relative to the ground state near equilibrium (2.6 a$_0$).
               The states shown are for the transitions connecting the 
               X $^3\Sigma^{-} \rightarrow  2~^{3}\Sigma^{-}$, 3$~^{3}\Sigma^-,$ A$~^3\Pi$, 2$~^{3}\Pi$, 3$~^{3}\Pi$ 
               states involved in the photodissociation process.
               (b) Dipole transition moments $D(R)$ (a.u.) for the ${\rm X}~^{3}\Sigma^{-} \rightarrow
 		{\rm A}~ ^{3}\Pi,  2~^{3}\Sigma^{-},   3~^{3}\Sigma^-, 2~^{3}\Pi, 2~^{3}\Pi, $ transitions.  The  MRCI + Q
		approximation with an AV6Z basis set  
		 was used to calculate the transition dipole moments.}
\label{fig1}
\end{figure*}
When the symmetry is reduced from C$_{\infty v}$ to C$_{2v}$, the correlating relationships
are $\sigma \rightarrow a_1$, $\pi \rightarrow$ ($b_1$, $b_2$) , $\delta \rightarrow$ ($a_1$, $a_2$). 
The active space consists of 16 electrons and 10 molecular orbitals ($6a_1$, $2b_1$, $2b_2$, $0a_2$) 
i.e., a (6, 2, 2, 0) active space.
To take account of short-range interactions we employed the multi-configuration-self-consistent-field (MCSCF) 
method~\cite{Werner1985,Knowles1985} available within the MOLPRO~\cite{Werner2010} suite of codes.  
This  procedure simultaneously optimises molecular orbitals and coefficients providing a truer representation of the molecular states.  
The averaging process was carried out on the lowest four $^3\Pi$, three $^3\Sigma^-$ and three $^3\Delta$ molecular states.

In Fig. \ref{fig1}(a) we illustrate all the triplet states involved in the photodissociation processes.
 All the potential energies in Fig. \ref{fig1} are given in eV relative to the ground-state
equilibrium bond distance $r_e$.  
Table 1 gives asymptotic properties of the molecular states.
Fig. \ref{fig1}(b) illustrates the dipole transition moments $D(R)$  as a function of internuclear separation $R$ 
connecting the X $\rm ^3\Sigma^-$  to the excited triplet  electronic states involved in the dynamics. 
 We note that the asymptotic separated-atom  energies in the MRCI+Q approximation 
 show suitable agreement with experimental atomic values (see Table 1).
 Our results show smooth transition dipole moments connecting all of the adiabatic 
 triplet states which leads us to conclude that the sensitivity 
of the transitions will be primarily due to the RV wave function envelope. 
Resonance behaviour can be expected to be primarily due to potential scattering in the outgoing dissociation channel.

Beyond $R$ = 20~a$_0$, the potential energies of the $3~^3\Sigma^-$ and $3~^3\Pi$ excited states 
can be approximated by the long-range multipole expansion,
\begin{equation}
 V(R) = \frac{Q_M}{2R^{\mathrm{3}}}-\frac{\alpha_d}{2R^{\mathrm{4}}}	+	V(R \rightarrow +\infty),  \label{multipole}
\end{equation}
where $Q_M$ is the quadrupole moment and $\alpha_{d}$ is the  dipole polarizability of the neutral atom.
$Q_M$ for a $p^4$ configuration is $-\frac{4}{5}e<r^2>$ for a $\Sigma$ 
state and $\frac{2}{5}e<r^2>$ for a $\Pi$ state \cite{Gentry1977}, 
where the $<r^2>$ value for sulfur is 5.065255 a.u ($^3P$ state) \cite{Fisher1972,Fisher1973} and $Q_M=0$ for H.
The atomic values for the dipole polarizability adopted here are respectively, $\alpha$(S)=18.0 and  $\alpha$(H)=4.5 
from the early work of Miller and Bederson \cite{MB1977}.
The long-range expansion of the ground and other excited electronic 
states includes just the last 2 terms of the right hand side in (\ref{multipole}),
i.e. $Q_M =0$. 
At short range internuclear distances, we fitted the potentials with the form, 
\begin{equation}
 V = A e^{-BR} +  C,
\end{equation}
where $A$, $B$ and $C$ are fitting coefficients. A similar approach was used for 
the TDMs to extend to long and short range internuclear distances.  Since our TDM's data 
are computed on the grid ($R_{min}$, $R_{max}$), so for $R < R_{min}$ a quadratic form 
\begin{equation}
V=a R^2 + b R + c,
\end{equation}
was used to connect to the united atom limit.  Similarly for $R > R_{max}$ we used 
the functional form,
\begin{equation}
V=a' e^{-b' R} + c',
\end{equation}
to connect to the separated atom limit.  The reduced mass used in our work was for the most abundant species 
and taken from the compilation of Huber and Herzberg \cite{hub79}.
%
%
%
%
		%
		%
		%
\begin{table}
\centering
\caption{The asymptotic separated-atom (SA) and the united-atom (UA) limits for the molecular states involved
               in the SH$^+$ photodissociation process. Energies E (eV) are relative to the separated atom 
              for the ground state.}
\vspace{2mm}
\scriptsize
\begin{tabular}{llccccc}
\hline
Molecular 		&Separated-atom 			&E (eV)		& E (eV)	&$\Delta E$		&United-atom  Cl$^{+}$		& Cl$^+$ \\
State			&Atomic state				&Expt.$^a$	& Theory$^b$	& (eV)			&Atomic state			& TDM's$^c$\\
			&					& 		&		&			&				& (a.u.)\\  
\hline
X$~^3\Sigma^-$	& S$^+$($3s^23p^3~^4S^o$) + H$(^2S)$		& 0.0		& 0.0		& -			& $3s^23p^4~^3P$		&\, - \\
A$~^3\Pi$ 		& S$^+$($3s^23p^3~^2D^o$) + H$(^2S)$ 	& 1.8439 	& 1.8066	&+0.0373		& $3s^23p^4~^3P$ 		& 0.0 \\
$2~^3\Sigma^-$	& S$^+$($3s^23p^3~^2D^o$) + H$(^2S)$ 		&1.8439 	& 1.8177	&+0.0262		& $3s^23p^3(^4S^o)4s~^3S^o$ 	&\, unknown\\
$2~^3\Pi$         	& S$^+$($3s^23p^3~^2P^o$) + H$(^2S)$   	&3.0445      	& 3.0150        &+0.0295        	& $3s3p^5~^3P^o$    		&\, 0.459\\
$3~^3\Sigma^-$   & S($3s^23p^4~^3P$) + H$^+(^1S)$      		&3.2384      	& 3.3321        &-0.0937        	& $3s^23p^3(^4S^o)3d~^3D^o$    	 &\, 0.997\\
$3~^3\Pi$        	 & S($3s^23p^4~^3P$) + H$^+(^1S)$       &3.2384   	& 3.3569        &-0.1185        	& $3s^23p^3(^4S^o)3d~^3D^o$     &\, 0.997\\
\hline
\end{tabular}
\begin{flushleft}
$^a$Energy values deduced from the NIST Atomic Spectra Database tabulations \cite{NIST2015}.\\
$^b$Energy values from MOLPRO \cite{Werner2010}, MRCI+Q  approximation (AV6Z basis).\\ 
$^c$Transition dipole moments (TDM's) with initial states, deduced from \cite{NIST2015,Tayal2004}.\\
\end{flushleft}
\end{table}

\normalsize

\subsection{Photodissociation Approach}

The direct photodissociation rate due to absorption from the initial RV level is
\begin{equation}
k_{fi}^{pd} = \int \sigma^{fi}(\lambda)I(\lambda)d\lambda s^{-1},
\end{equation}
where $\sigma^{fi}(\lambda)$ is the direct photodissociation cross section for absorption 
into the RV continuum of electronic state $f$ and $I$ is the mean intensity of the radiation field.  
Given that the radiation field in most environments is unknown, we focus in this 
work on the cross section, which for a bound-free transition from initial
state $i$ to final state $f$ for an electric dipole transition, can be expressed as \cite{rau02}
\begin{equation}
\sigma^{fi}(E_{\rm ph})=\frac{2\pi^2 e^{\mathrm{2}} \hbar}{m_ec}\frac{df}{dE_{\rm ph}},
\label{sig1}
\end{equation}
where $m_e$ is the mass of the electron, $E_{\rm ph}$ is the photon energy, 
and the other constants have their usual meanings.
Eq.~(\ref{sig1}) can be written as
\begin{equation}
\sigma^{fi}(E_{\rm ph})=\frac{4\pi^2}{3}\alpha E_{\rm ph}  |\langle \Phi_f(\vec{r},\vec{R}) |\vec{r} |
\Phi_i(\vec{r},\vec{R} \rangle|^2,
\label{sig2}
\end{equation}
when we used the continuum differential oscillator strength and the definition of the fine-structure constant,
$\alpha=e^2/\hbar c$.
The continuum differential oscillator strength in the length gauge is given as
\begin{equation}
\frac{df}{dE_{\rm ph}}=\frac{2m_e}{3\hbar^2} E_{\rm ph} |\langle \Phi_f(\vec{r},\vec{R}) |\vec{r} |
\Phi_i(\vec{r},\vec{R} \rangle|^2,
\label{dfde}
\end{equation}
where $\Phi(\vec{r},\vec{R})$ is the total molecular wave function, $\vec{r}$ is the electronic
coordinate vector, and $\vec{R}$ is the internuclear vector.
The numerical value of the pre-factor becomes $2.689\times 10^{-18}$ 
when all quantities in Eq.~(\ref{sig2}) are taken in atomic units, and
the cross section is expressed in cm$^2$.  By applying separation of variables to the
electronic and nuclear coordinates of $\Phi$,
the electric dipole transition moment function becomes
\begin{equation}
D^{fi}(R)=\langle \phi_f(\vec{r}|R) | \vec{r} |\phi_i(\vec{r}|R) \rangle,
\label{dipole}
\end{equation}
where $\phi(\vec{r}|R)$ is the electronic molecular wave function 
for fixed $R$ and integration is taken over all electronic coordinates.
Then, the photodissociation cross section from initial rovibrational 
level $v''J''$ can be expressed as \cite{1988kirby}
\begin{eqnarray}
\sigma^{fi}_{v''J''}(E_{\rm ph})=&2.689 \times 10^{\mathrm{-18}}
E_{\rm ph}  ~g~ & \nonumber \\
 &\times \sum_{J'}  \left ( \frac{1}{2J''+1}
S_{J'}|D^{fi}_{k'J',v''J''}|^{\mathrm{2}} \right )
\mathrm{cm}^{\mathrm{2}},&
\end{eqnarray}
where the H\"{o}nl-London factors, $S_{J'}(J'')$ for
a ${\Sigma \leftarrow \Sigma}$
electronic transition are given by
\begin{eqnarray}
S_{J'}(J'')&=&
\biggl \{
\begin{array}{ll}
J''-1,&~J'=J''-1~(\mbox{P-branch})\\
J''+1,&~J'=J''+1~(\mbox{R-branch}),
\end{array} 
\end{eqnarray}
and for a
${\Pi \leftarrow \Sigma}$
electronic transition are written as                                                                                                                                                                                                                                                                   
\begin{eqnarray}
S_{J'}(J'')&=&
\Biggl \{
\begin{array}{ll}
(J''-1)/2,&~J'=J''-1~(\mbox{P-branch})\\
(2J"+1)/2,&~J'=J"~(\mbox{Q-branch})\\
(J''+1)/2,&~J'=J''+1~(\mbox{R-branch}).
\end{array} 
\end{eqnarray}
The matrix element of the electric dipole transition moment for absorption from the rovibrational level $v''J''$ in state
$i$ to the continuum $k'J'$ in state $f$ is $D^{fi}_{k'J',v''J''}=\langle \chi_{k'J'}(R)|D^{fi}(R)|\chi_{v''J''}(R)\rangle$
where $J$ is the angular momentum of nuclear motion and $g$ is the degeneracy factor, with the integration taken over $R$.
The degeneracy factor is written as
\begin{equation}
g=\frac{2-\delta_{\mathrm{0,\Lambda'+\Lambda''}}}{2-\delta_{\mathrm{0,\Lambda''}}},
\end{equation}
where $\Lambda'$ and $\Lambda''$ are the angular momenta projected along the
nuclear axis for the final and initial electronic states.
The bound and continuum rovibrational wave functions, $\chi_{v''J''}(R)$ and $\chi_{k'J'}(R)$
are solutions of the radial Schr$\ddot{{\rm o}}$dinger equation for nuclear
motion on the $i$ and  $f$ state potential curves.
The continuum wave functions are normalized such that
they behave asymptotically as
\begin{equation}
\chi_{k'J'}(R) = \biggl (\frac{2\mu}{\pi^2E^\prime} \biggr )^{1/4} \sin(k'R-\frac{\pi}{2}J'+\eta_{J'}),
\end{equation}
where $\mu$ is the molecular reduced mass \cite{hub79}, $\eta_{J'}$ the phase shift, $E^\prime$ the center-of-mass kinetic energy for the
free particles, and the wave vector is given by $k^\prime = \sqrt{2\mu E^\prime}$.
The wave functions were obtained numerically using the standard Numerov method \cite{bla67,coo61,joh77}
with a step size of 0.001 a$_0$ over internuclear distances
$0.1<R<200$~a$_{\mathrm{0}}$.

If the density of the environment is sufficiently high, then the populations 
of the $v^{\prime\prime}J^{\prime\prime}$ levels may be assumed
to be thermalized and it is appropriate to consider the so-called local thermodynamic equilibrium (LTE) cross section,
\begin{eqnarray}
\sigma(\lambda,T)=\frac{1}{Q(T)} 
\sum_{v''}\sum_{J''}(2J''+1)& \exp{[-(E_{v^{\prime\prime}=0,J^{\prime\prime}=0}-E_{v^{\prime\prime}J^{\prime\prime}})/k_{\mathrm{b}}T]} &  \nonumber \\
 &\times \sigma_{v^{\prime\prime}J^{\prime\prime}}(\lambda),&
\label{siglte}
\end{eqnarray}
\cite{arg74}, which depends only on the gas temperature $T$ and the photon wavelength $\lambda$ where 
$k_{\rm b}$ is Boltzmann's constant, with the RV partition function given by 
\begin{equation}
Q(T)=\sum_{v^{\prime\prime}}\sum_{J''}(2J''+1)
\exp{[-(E_{v''=0,J''=0}-E_{v''J''})/k_{\mathrm{b}}T]}.
\label{part}
\end{equation}

\section{Results and Discussion}\label{secResults}

\subsection{Spectroscopic Properties}
  
 Using the ab initio data for the X$~^3\Sigma^-$ potential described above,  we
 obtained the spectroscopic constants given in Table~\ref{tab2}. This is compared
 to experimental data \cite{hub79,dun79} and the previous theoretical results of Khadri et al. \cite{kha06} and found
 to be in reasonable agreement. The X$~^3\Sigma^-$ is found to support 22 vibrational
 levels, with a maximum rotational level $J''=62$, for a total of 841 rovibrational levels.
	%
  	%
\begin{table}
\caption{Spectroscopic constants for the X~$^3\Sigma^-$ electronic ground state compared to
		available experimental and theoretical data.\label{tab2}}
\begin{indented}
\item[]\begin{tabular}{@{}*{7}{cccccc}}
\hline
$r_e$ 		& $D_e$ 		& $D^0_0$		 & $\omega_e$ 		& $\omega_ex_e$		 & Notes \\
(a$_0$)	 & (eV) 		& (eV)			& (cm$^{-1}$) 		& (cm$^{-1}$) 		& \\ 
\hline
2.56 		&  3.843 		& 3.684		 & 2601.1 			& 54.1 			& This work\\
2.577      	& -         	 	&  -        		& 2558.0			 & 0				 & Ref.  \cite{dun79} \\
2.597    	&  -         		& 3.48  		 & 2547.7 			& 49.3 			& Ref. \cite{hub79} \\
2.572     	 & -         		 & 3.488 		& 2555.2			 & 49.0 			& Ref. \cite{kha06} \\
\hline
\end{tabular}
\end{indented}
\end{table}  
  
 \subsection{Partial cross sections for rovibrationally-resolved photodissociation}
\begin{figure}
\includegraphics[scale=0.5]{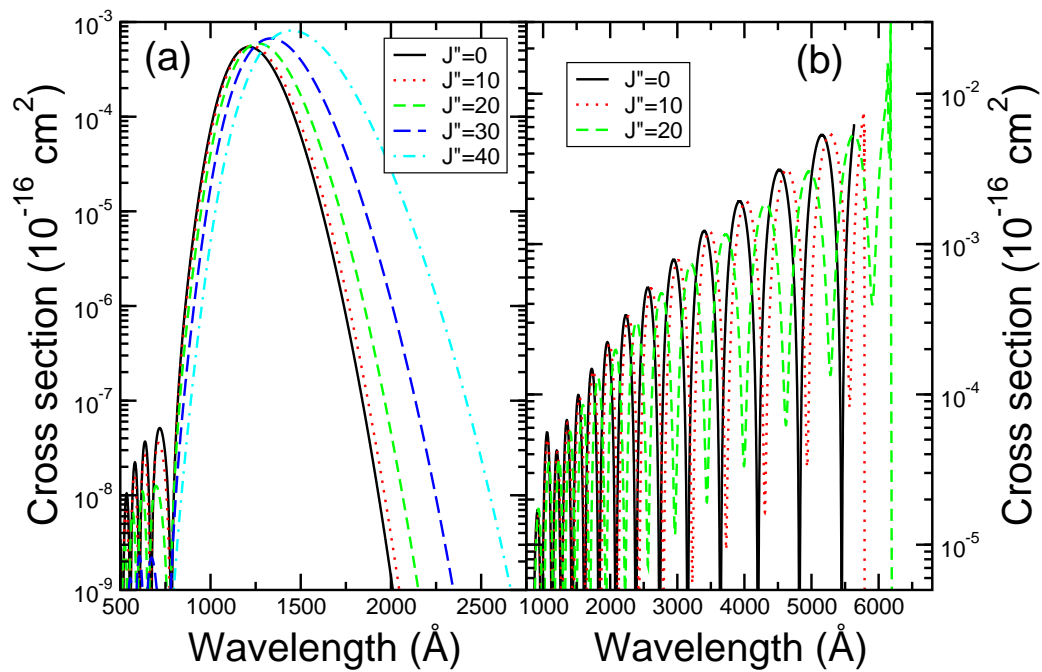}
\caption{The computed partial SH$^+$ photodissociation cross-sections for 
the 2~$^{3}\Sigma^{-} \leftarrow$ X$~^{3}\Sigma^{-}$ rovibrational transition 
for (a) $v^{\prime\prime}=0$ and select $J^{\prime\prime}$ and (b) $v^{\prime\prime}=15$ and select $J^{\prime\prime}$.}
\label{fig2}
\end{figure}

\begin{figure}
\includegraphics[scale=0.5]{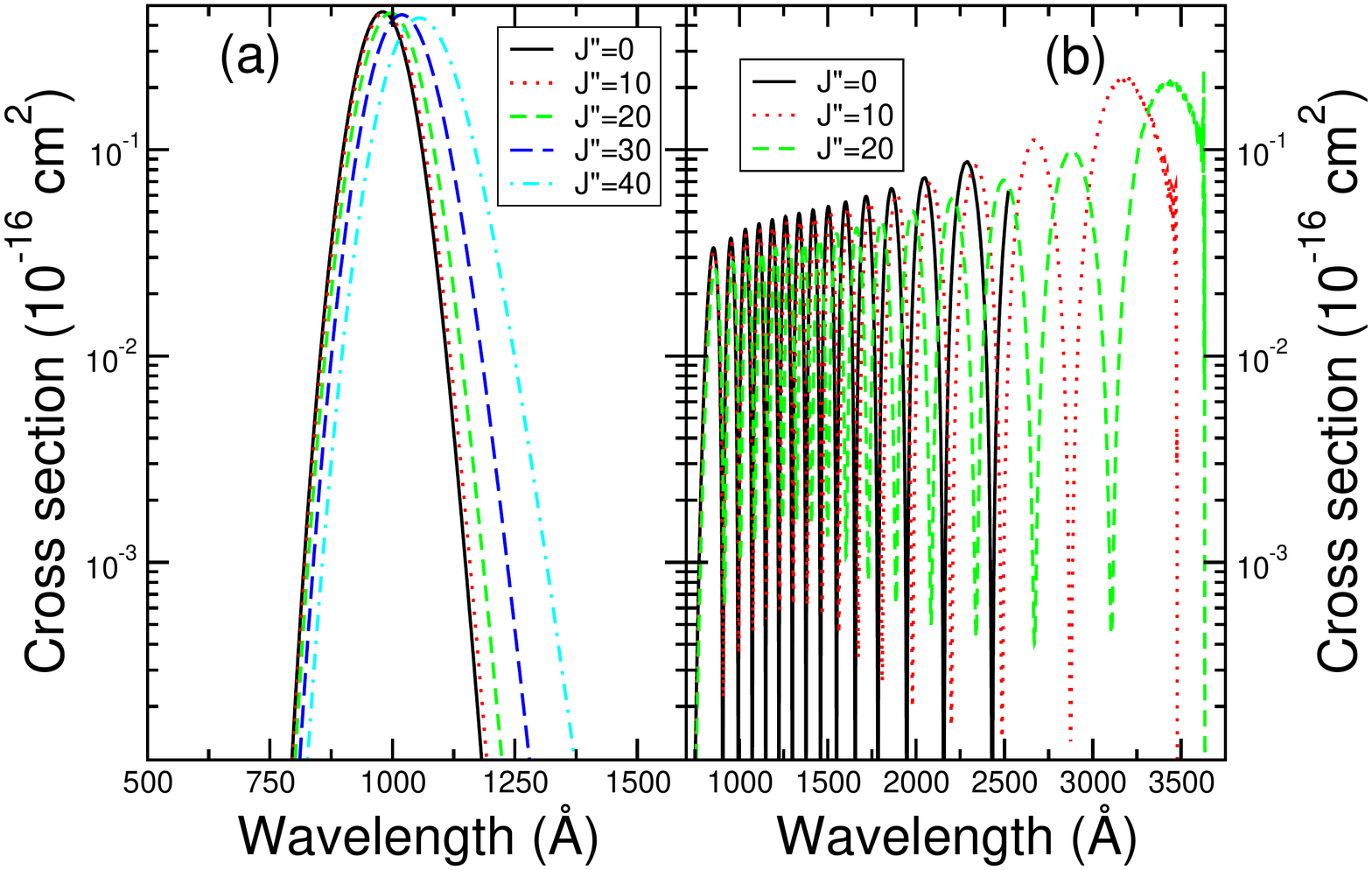}
\caption{The computed partial SH\textsuperscript{+} photodissociation cross-sections 
for the $3$ $^{3}\Sigma ^{-}$ $\leftarrow$ $X$ $^{3}\Sigma^{-}$ rovibrational 
transition for (a) $v^{\prime\prime}=0$ and select $J^{\prime\prime}$ and (b) $v^{\prime\prime}=15$ and select $J^{\prime\prime}$.
\label{fig3}}
\end{figure}

\begin{figure}
\includegraphics[scale=0.5]{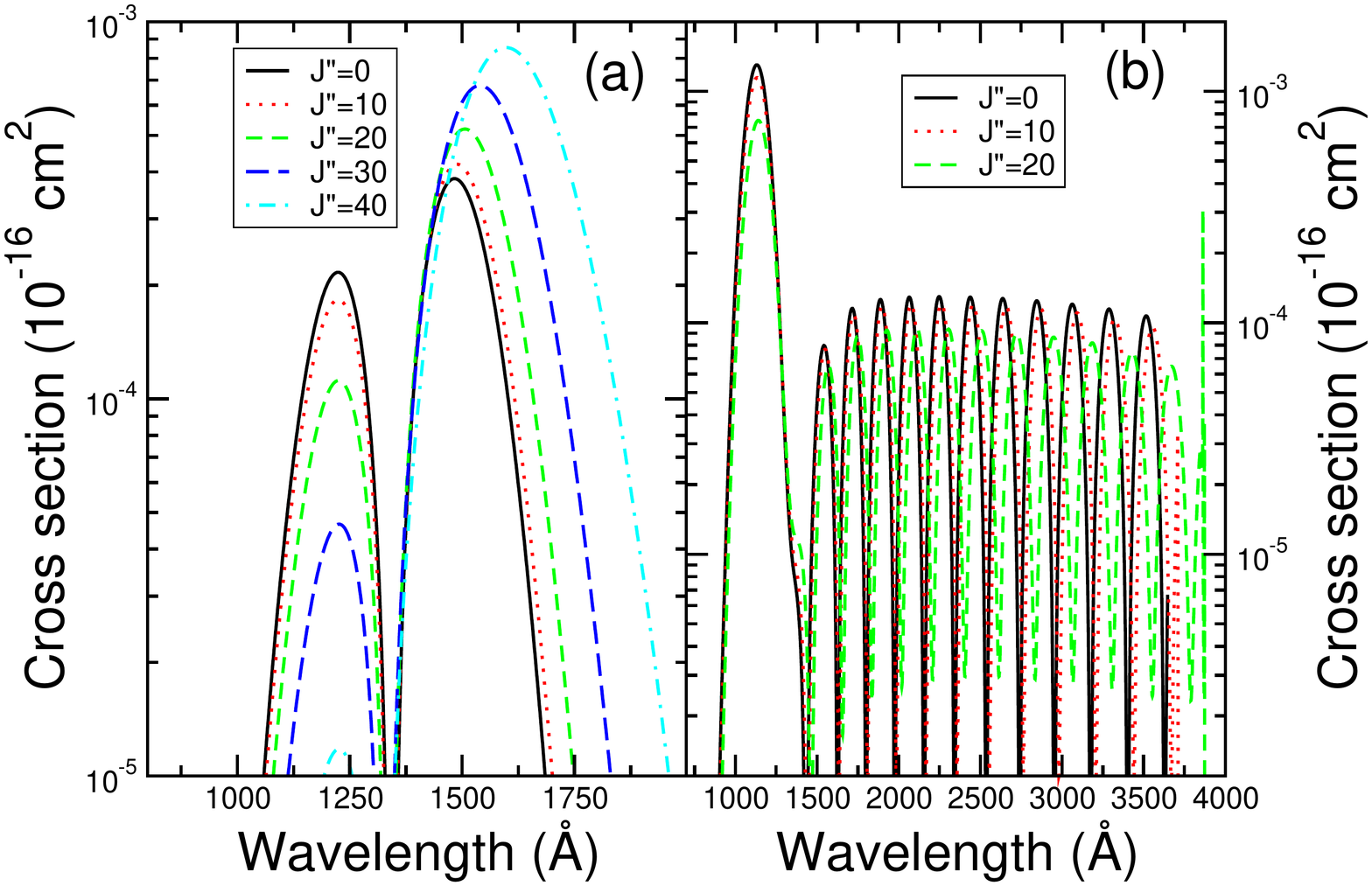}
\caption{The computed partial SH\textsuperscript{+} photodissociation 
cross-sections for the $2$ $^{3}\Pi$ $\leftarrow$ X~$^{3}\Sigma^{-}$ rovibrational 
transition for (a) $v^{\prime\prime}=0$ and select $J^{\prime\prime}$ and (b) $v^{\prime\prime}=15$ and select $J^{\prime\prime}$.
\label{fig4}}
\end{figure}

\begin{figure}
\includegraphics[scale=0.5]{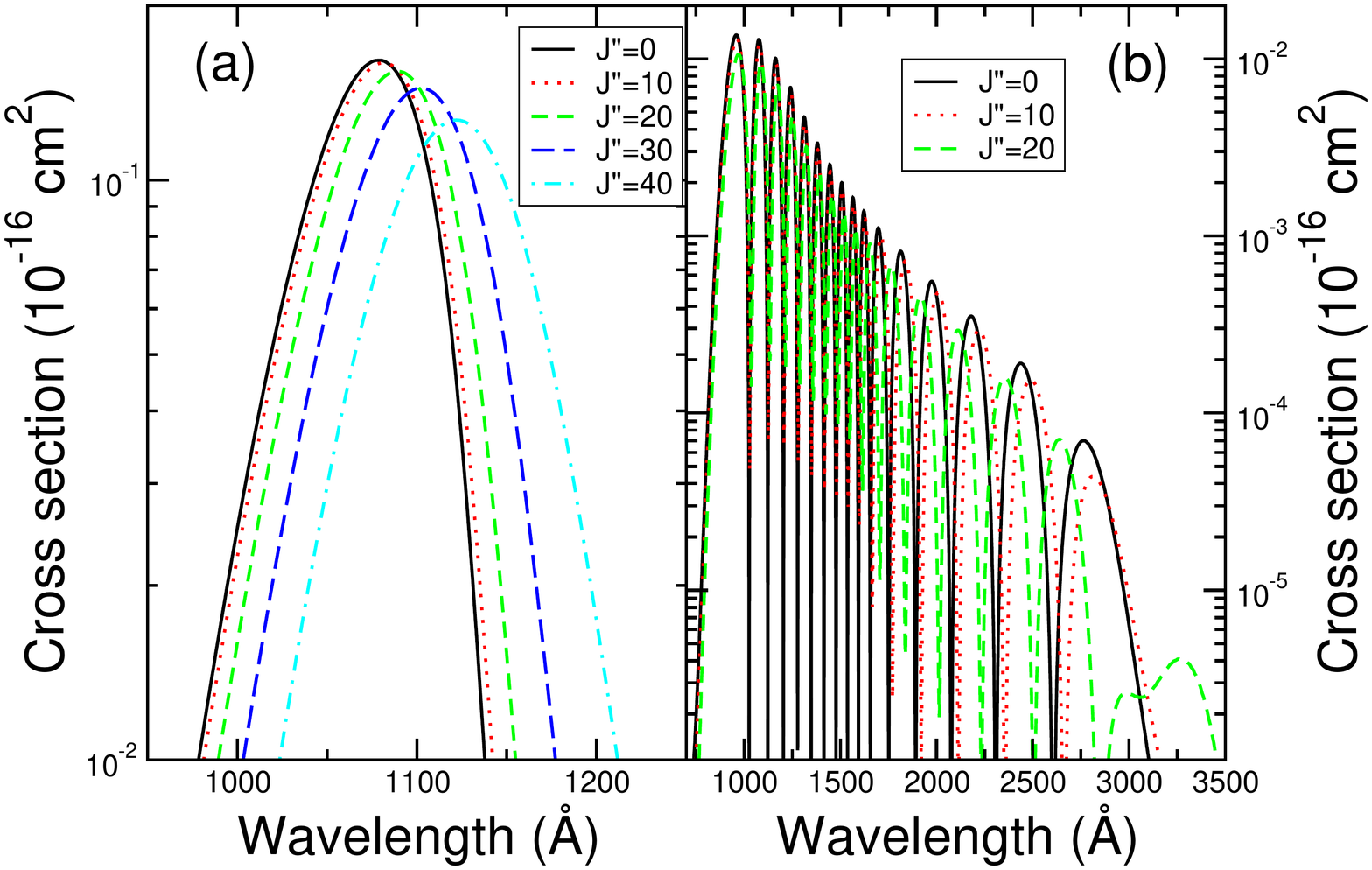}
\caption{The computed partial SH\textsuperscript{+} photodissociation 
cross-sections for the $3$ $^{3}\Pi$ $\leftarrow$ X~$^{3}\Sigma^{-}$ rovibrational 
transition for (a) $v^{\prime\prime}=0$ and select $J^{\prime\prime}$ and (b) $v^{\prime\prime}=15$ and select $J^{\prime\prime}$.
\label{fig5}}
\end{figure}

\begin{figure}
\includegraphics[scale=0.5]{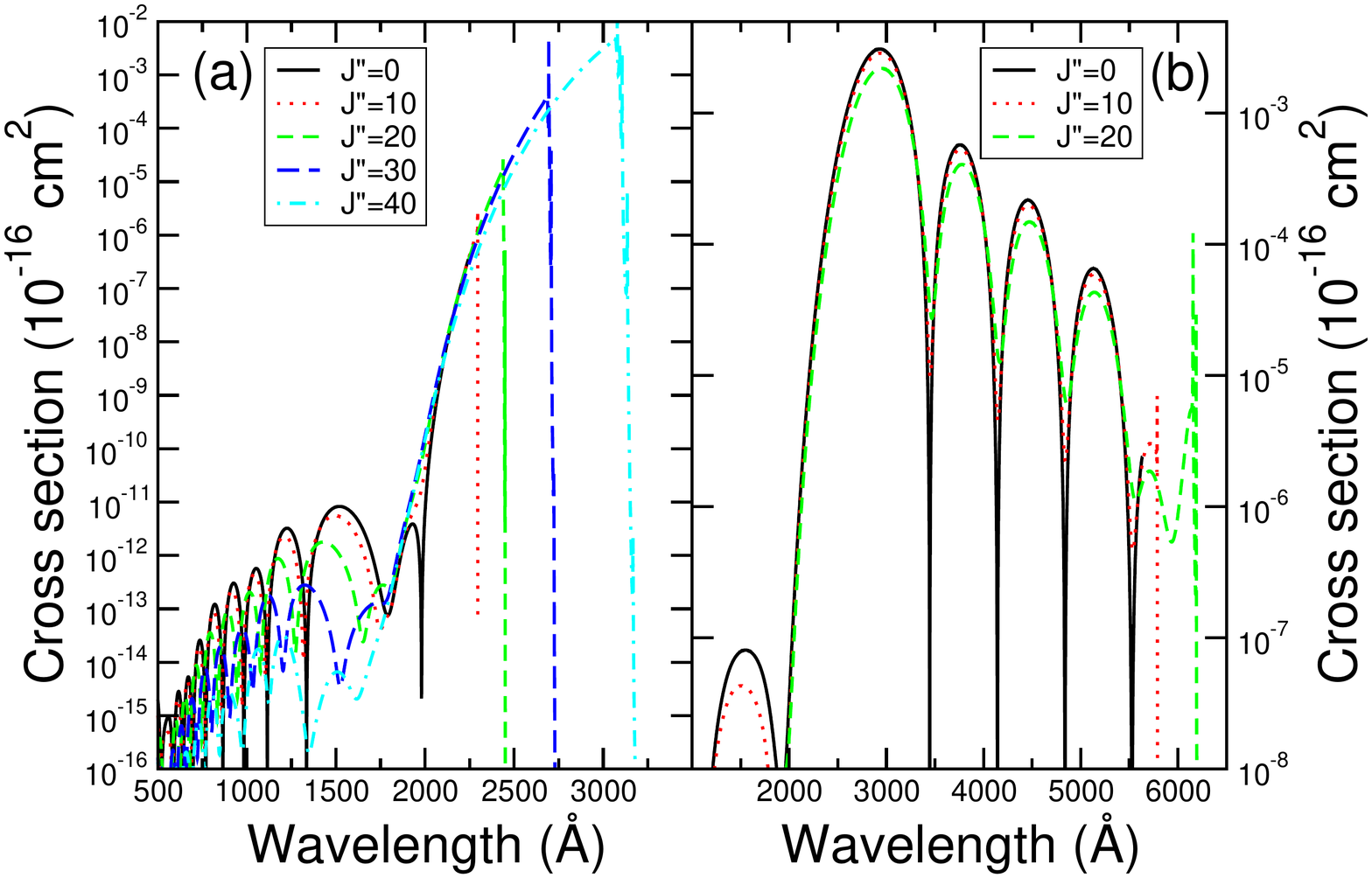}
\caption{The computed partial SH\textsuperscript{+} photodissociation cross-sections 
for the A $^{3}\Pi$ $\leftarrow$ X~$^{3}\Sigma^{-}$ rovibrational transition 
for (a) $v^{\prime\prime}=0$ and select $J^{\prime\prime}$ and (b) $v^{\prime\prime}=15$ and select $J^{\prime\prime}$.
\label{fig6}}
\end{figure}

 A selection of the partial cross sections $\sigma_{v''J''}$ for the 
$2$ $^{3}\Sigma ^{-}$ $\leftarrow$ $X$ $^{3}\Sigma^{-}$ transition 
as a function of the photon wavelength are presented in Figure~\ref{fig2} 
 with results for $v^{\prime\prime}=0$ and select $J^{\prime\prime}$ given
 in Figure~\ref{fig2}a  and    $v^{\prime\prime}=15$ in Figure~\ref{fig2}b. As  expected the cross sections
 migrate to larger photon energies due to the decreasing photon threshold energy 
with increasing $v^{\prime\prime}$ and/or $J^{\prime\prime}$. 
Orbiting resonances due to quasi-bound levels in the upper state
 become evident near threshold for relatively large $v^{\prime\prime}$ and/or $J^{\prime\prime}$.  
The same trends are evident for the other four electronic transitions give in Figures~\ref{fig3}-\ref{fig6}.
Of note, is the  A~$^{3}\Pi$ $\leftarrow$ $X$ $^{3}\Sigma^{-}$ transition, which has a very
 weak cross from $v^{\prime\prime} =0$, $J^{\prime\prime} =0$ due to the fact the most of the oscillator strength
resides in bound-bound transitions as a opposed to the bound-free case. 
However,  the cross sections become significant, comparable to
other electronic transitions, for large $v^{\prime\prime}$ and/or $J^{\prime\prime}$, 
but also at wavelengths longer than ~$\sim$2000~\AA.

\begin{figure}
\includegraphics[scale=0.5]{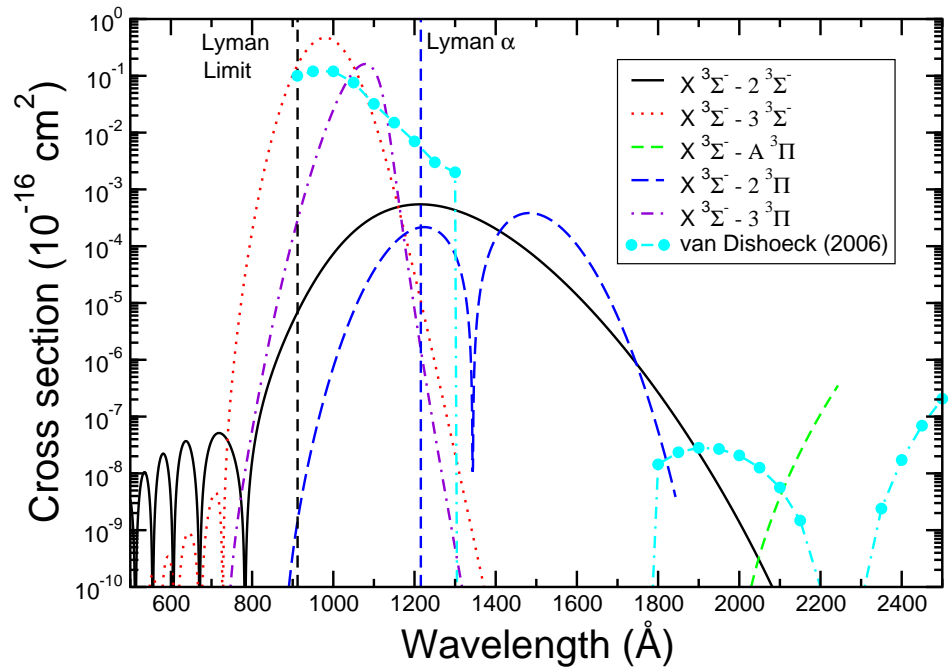}
\caption{Comparison of SH\textsuperscript{+} photodissociation cross sections 
for $v^{\prime\prime}=0$ and $J^{\prime\prime}=0$ with estimates from Ref. \cite{van06}.
\label{fig7}}
\end{figure}

Figure~\ref{fig7} compares the cross sections for each transition from the electronic 
and rovibrational ground-state $v^{\prime\prime} =0$, $J^{\prime\prime} =0$ which is relevant for cold to warm interstellar
diffuse and translucent clouds. The 3 $^{3}\Sigma ^{-}$ $\leftarrow$ $X$ $^{3}\Sigma^{-}$ 
and 3 $^{3}\Pi$ $\leftarrow$ $X$ $^{3}\Sigma^{-}$ dominate with most of their cross section falling between
the H Lyman photoionization limit and the H Lyman$\alpha$ being the region were the ISM 
radiation field is most intense. The other transitions are at least two orders of magnitude weaker with their
peaks lying at longer wavelengths.  

As far as we are aware, the current work is the first explicit photodissociation calculations 
for the SH$^+$ radical ion. An estimate was made in van Dishoeck et al. \cite{van06} of the SH$^+$ cross section
by scaling that of CH$^+$. As illustrated in Figure~\ref{fig7} , there is suitable agreement, 
however the current results are about a factor of 3 larger, therefore we would expect the photodissociation rate
to be enhanced by a similar amount. 

\subsection{LTE photodissociation cross sections}\label{LTE}

\begin{figure}
\includegraphics[scale=0.5]{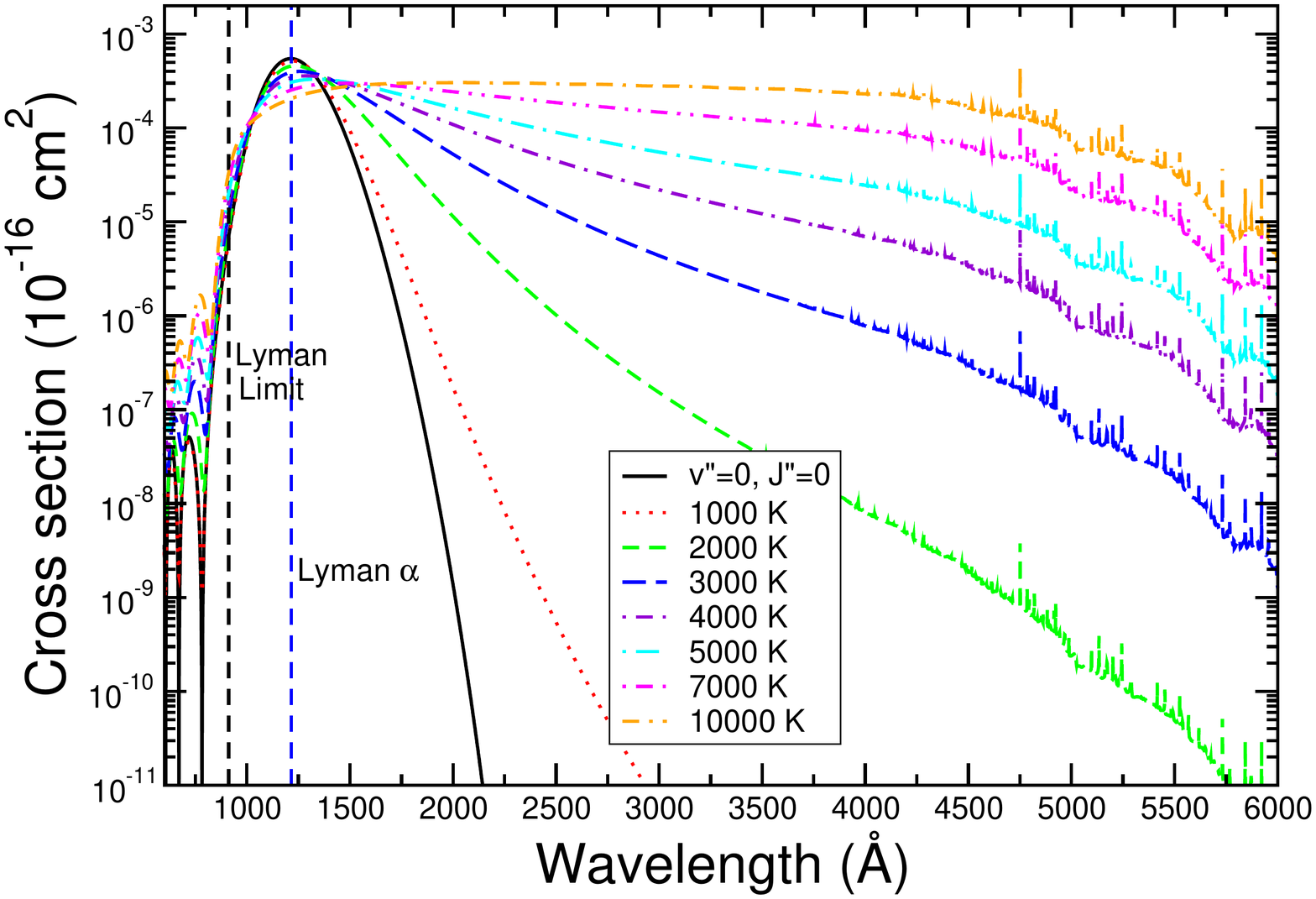}
\caption{Total SH$^{+}$ 2$~^3\Sigma^- \leftarrow
\mathrm{X}~^3 \Sigma^{\mathrm{-}}$ LTE photodissociation cross section for temperatures
from 1000 to 10,000 K. The $v''=0$, $J''=0$ partial
cross section is also plotted for comparison.
\label{fig8}}
\end{figure}

\begin{figure}
\includegraphics[scale=0.5]{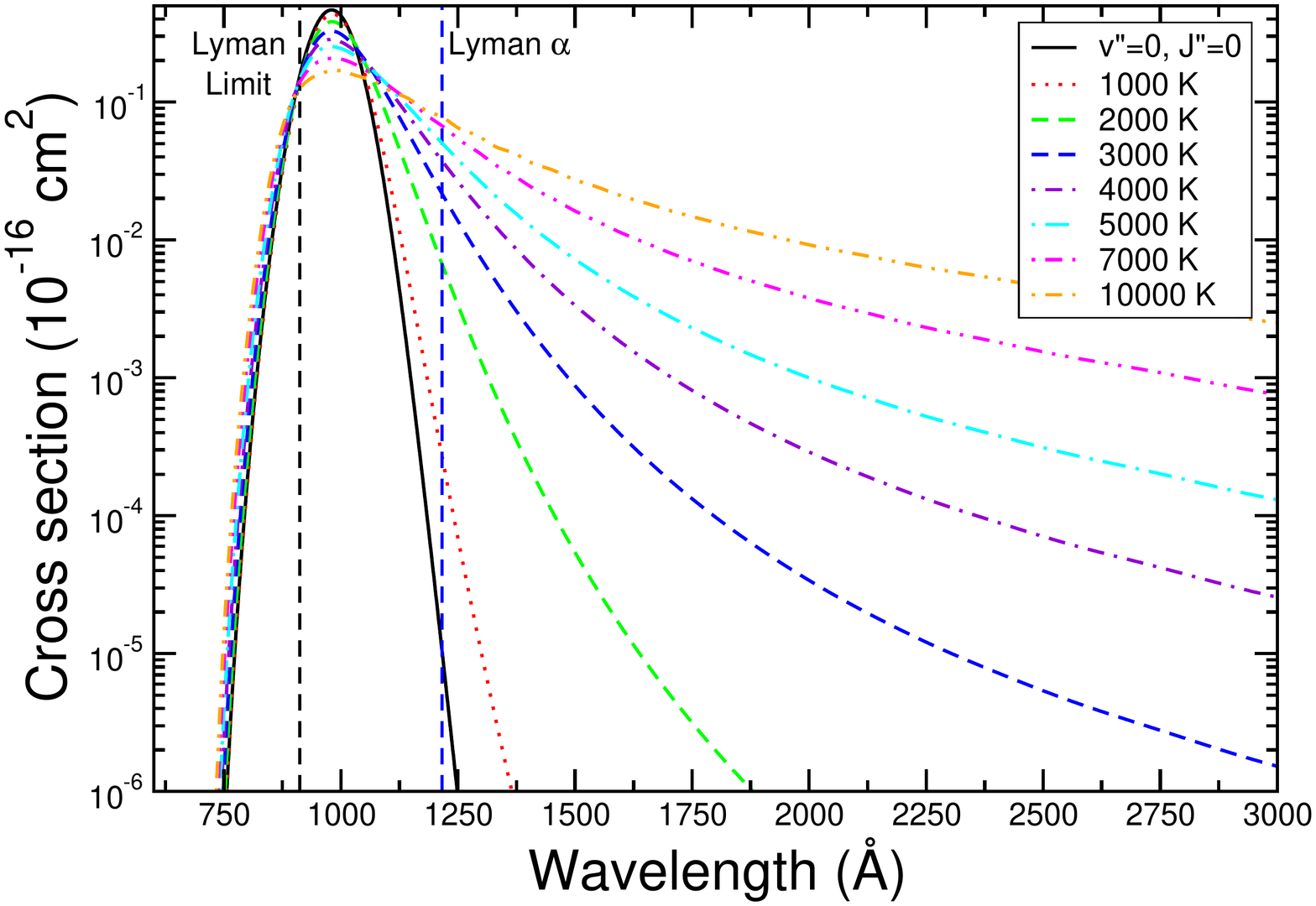}
\caption{Total SH$^{+}$ 3$~^3\Sigma^- \leftarrow
\mathrm{X}~^3 \Sigma^{\mathrm{-}}$ LTE photodissociation cross section for temperatures
from 1000 to 10,000 K. The $v''=0$, $J''=0$ partial
cross section is also plotted for comparison.
\label{fig9}}
\end{figure}

\begin{figure}
\includegraphics[scale=0.5]{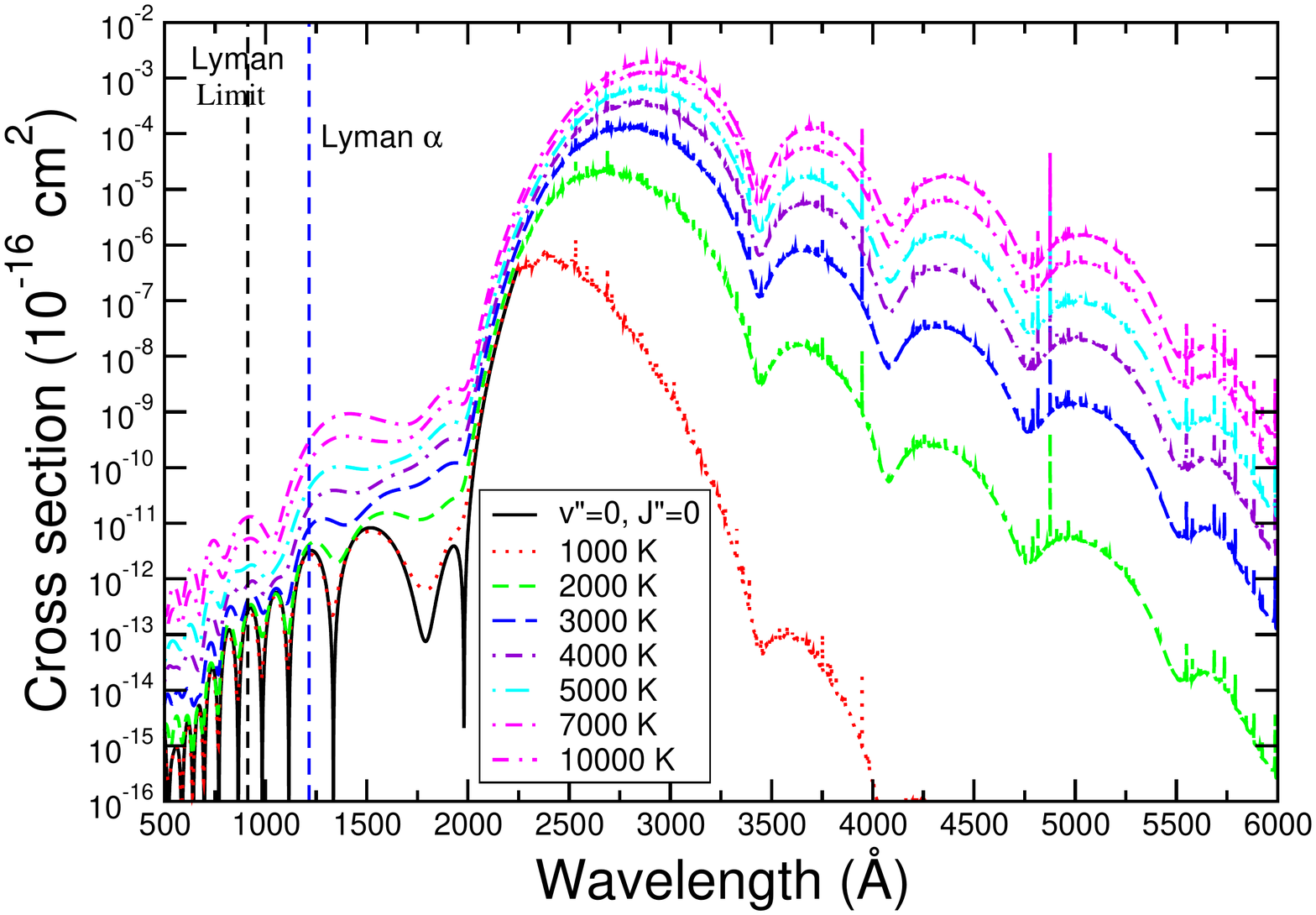}
\caption{Total SH$^{+}$ A$~^3\Pi \leftarrow
\mathrm{X}~^3 \Sigma^{\mathrm{-}}$ LTE photodissociation cross section for temperatures
from 1000 to 10,000 K. The $v''=0$, $J''=0$ partial
cross section is also plotted for comparison.
\label{fig10}}
\end{figure}

\begin{figure}
\includegraphics[scale=0.5]{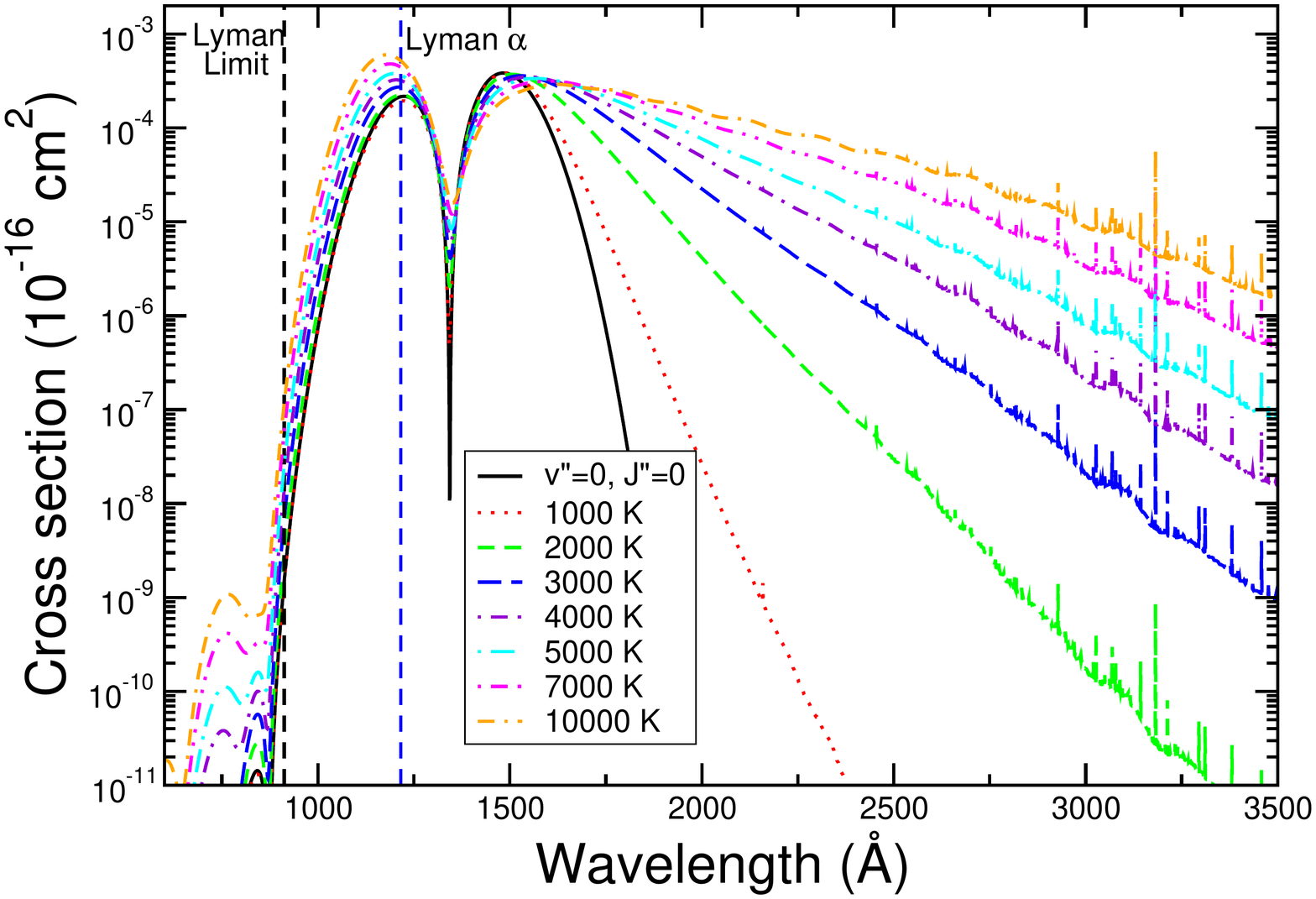}
\caption{Total SH$^{+}$ 2$~^3\Pi \leftarrow
\mathrm{X}~^3 \Sigma^{\mathrm{-}}$ LTE photodissociation cross section for temperatures
from 1000 to 10,000 K. The $v''=0$, $J''=0$ partial
cross section is also plotted for comparison.
\label{fig11}}
\end{figure}

\begin{figure}
\includegraphics[scale=0.5]{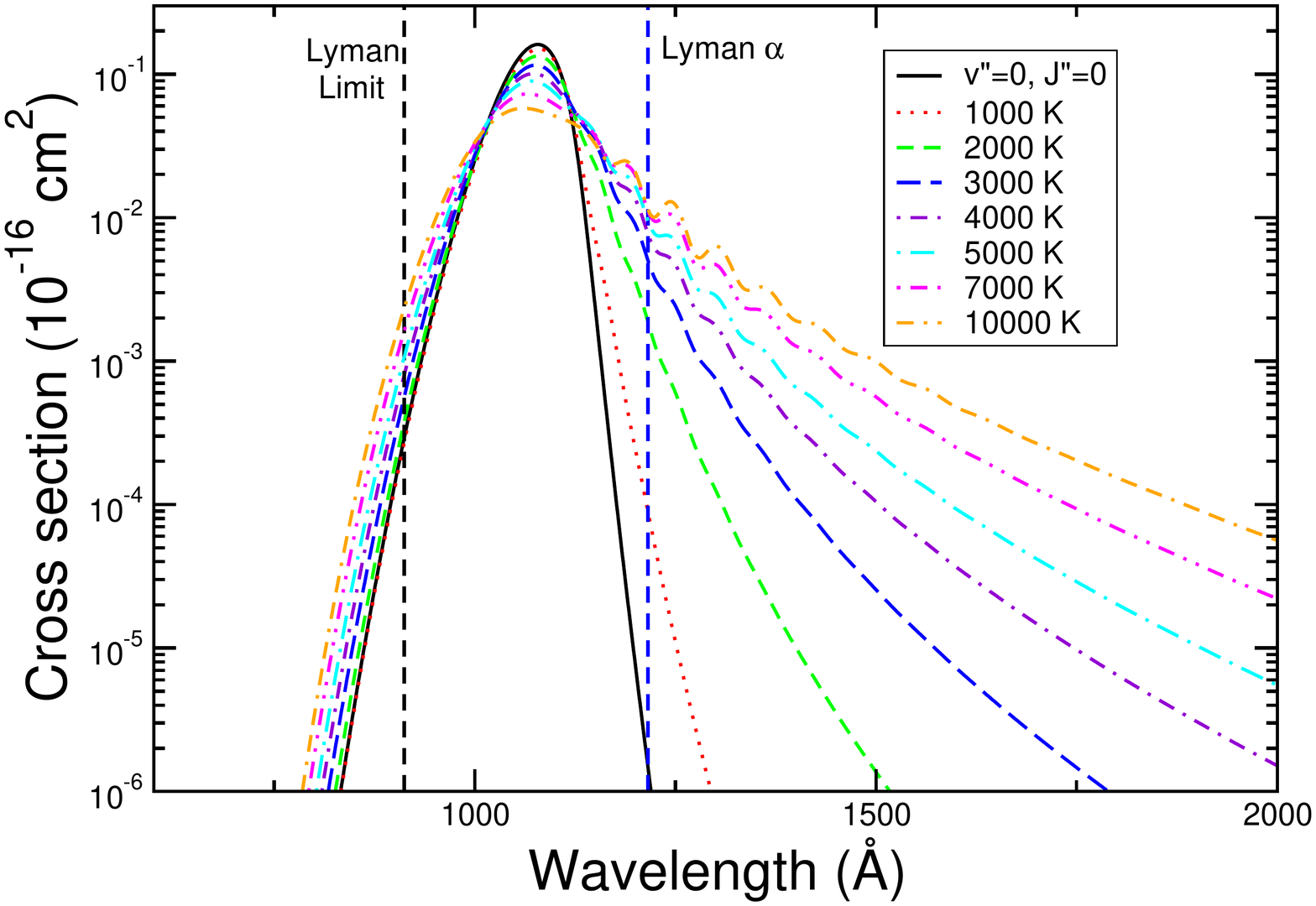}
\caption{Total SH$^{+}$ 3$~^3\Pi \leftarrow
\mathrm{X}~^3 \Sigma^{\mathrm{-}}$ LTE photodissociation cross section for temperatures
from 1000 to 10,000 K. The $v''=0$, $J''=0$ partial
cross section is also plotted for comparison.
\label{fig12}}
\end{figure}

Figures~\ref{fig8}-\ref{fig12} present the LTE cross sections, which assume a thermal 
population for the RV levels in X~$^3\Sigma^-$, for all five transitions. 
The LTE cross sections are shown for plasmas temperatures in the range from 
1000 to 10,000 K and indicate  two main features: i) cross sections at longer 
wavelength increase with increasing temperature and ii) cross sections
at the peak for $v^{\prime\prime} =0$, $J^{\prime\prime} =0$ typically 
decrease as shown in Figure~\ref{fig8}. However, the overall difference between the RV ground state cross section
and that for 1000 K is not large.

In Figure~\ref{fig13}, the LTE cross sections for all five transitions are compared at 3000 K. 
This should be compared to Figure~\ref{fig7} for  $v^{\prime\prime} =0$, $J^{\prime\prime} =0$
where it is seen that the cross sections are larger in the LTE case for wavelengths longer than $\sim$1500 \AA. 
In particular, the   A~$^{3}\Pi$ $\leftarrow$ $X$ $^{3}\Sigma^{-}$ transition,
which is negligible for $v^{\prime\prime} =0$, $J^{\prime\prime} =0$ results 
in a large LTE cross section peak, dominating all other transitions, near 2800~\AA.


\section{Astrophysical Applications}
The SH$^+$ radical ion, sulfanylium, was not detected in the ISM until as late as 2010 \cite{ben10}. 
It is however, an important tracer of gas condensations in dense regions and also probes
the warm surface layers of PDRs \cite{nag13}. Furthermore, its abundance is expected to be enhanced
 in X-ray dominated regions (XDRs) \cite{abe08}. In their model of the Orion Bar PDR,
Nagy et al. \cite{nag13} find that photodissociation accounts for a maximum of about 4.4\% of
 the total destruction rate of SH$^+$, since reactive collisions with H and dissociative recombination by
electrons are more efficient. However, they adopted the estimated cross section 
of Ref. \cite{van06} for $v^{\prime\prime} =0$, $J^{\prime\prime} =0$. 
 We point out that the adoption of the current
cross sections would enhance the photodissociation contribution to greater than 10\%.
 We note that the photodissociation rates are not given here as they are sensitive to
the local radiation field and dust properties. The latter is quite different in the Orion Bar 
from the average ISM of the galaxy. The densities and temperatures (10$^5$-10$^6$ cm$^{-3}$ and $\sim$1000~K) 
of the Orion Bar PDR begin to approach the regime where photodissociation from 
excited states might contribute which is currently neglected in all models. Furthermore, LTE conditions are 
almost satisfied, but at 1000~K there is not a significant difference 
between the LTE and $v^{\prime\prime} =0$, $J^{\prime\prime} =0$ cross sections

\begin{figure}
\includegraphics[scale=0.5]{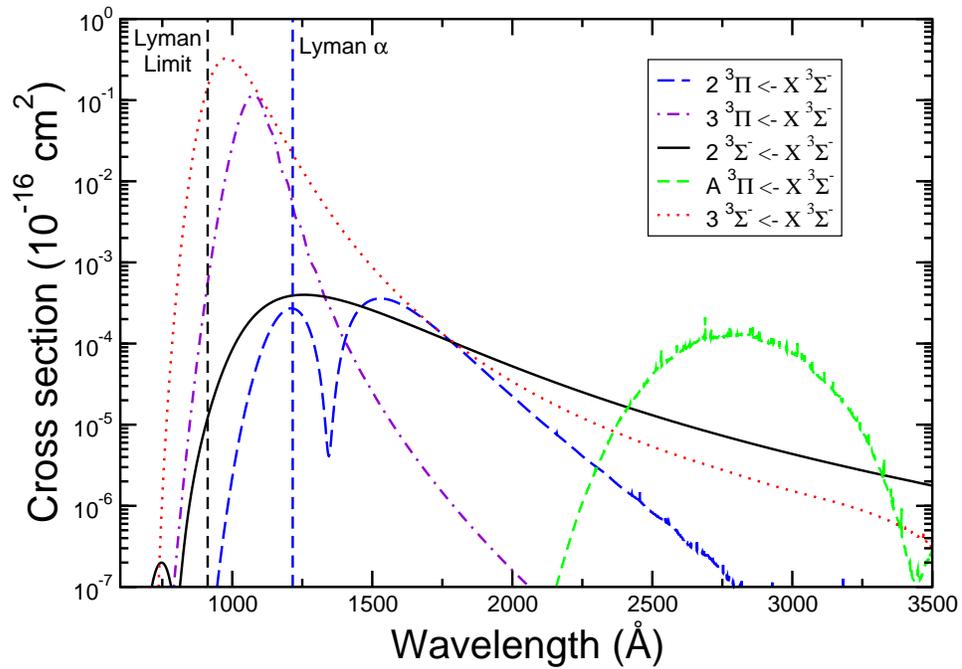}
\caption{Total SH$^{+}$ LTE photodissociation cross section at
3000 K for all electronic transitions. Blackbody radiation curves (dotted lines) for various radiation temperatures
are plotted for comparison (to be added).
\label{fig13}}
\end{figure}

For higher density and temperature environments, such as proto-planetary discs (PPDs) and planetary and 
stellar atmospheres, the LTE cross sections are more appropriate and molecular ion abundances
may be larger \cite{wir88,sta97}. We note that the LTE cross sections contribute to 
the bound-free opacity at longer wavelengths as illustrated in Figure~\ref{fig13}. To aid in astrophysical
models, all SH$^+$ photodissociation cross sections are posted on the
 UGA Molecular Opacity Project website\footnote{www.physast.uga.edu/ugamop}.

\section{Conclusions}\label{secConclusions}

Using ab initio potentials and transition dipole moment functions, accurate cross-section calculations 
have been performed for the photodissociation of SH\textsuperscript{+} through the transitions 
$2$ $^{3}\Sigma ^{-}$ $\leftarrow$ X~$^{3}\Sigma^{-}$, $3$ $^{3}\Sigma ^{-}$ $\leftarrow$ X $^{3}\Sigma^{-}$, 
A $^{3}\Pi$ $\leftarrow$ X $^{3}\Sigma^{-}$, $2$ $^{3}\Pi$ $\leftarrow$ X $^{3}\Sigma^{-}$, 
and $3$ $^{3}\Pi$ $\leftarrow$ X $^{3}\Sigma^{-}$.  The partial cross-sections have been evaluated 
from the vibrational levels $v^{\prime\prime}=0-19$ of the X $^{3}\Sigma^{-}$ electronic state for wavelengths from threshold to
500 \AA. Assuming a Boltzmann distribution of rovibrational levels in the ground electronic state, 
so-called LTE cross sections have also been computed for all five electronic transitions for 1000-10,000 K.
While it is found the 3 $^{3}\Sigma^{-} \leftarrow$ X~$^{3}\Sigma^{-}$ and 
3 $^{3}\Pi$ $\leftarrow$ X $^{3}\Sigma^{-}$ transitions give the 
dominant cross sections near the Lyman limit, important contributions are
provided by the A~$^{3}\Pi$ $\leftarrow$ X $^{3}\Sigma^{-}$  for high rovibrational
 levels and, in the case of LTE, high temperatures particularly for longer wavelengths greater $\sim$2500 \AA.

\ack
ECMcM acknowledges support from the UGA Center for Undergraduate Research Opportunities. 
GS acknowledges travel support by the International Cooperation and Exchange 
Foundation of CAEP.  PCS acknowledges partial support from HST-AR-11776.01-A  which was provided by NASA through
a grant from the Space Telescope Science Institute, which is operated
by the Association of Universities for Research in Astronomy, Incorporated,
under NASA contract NAS5-26555.  
The hospitality of the University of Georgia at Athens (UGA) is gratefully acknowledged by B MMcL during 
recent research visits.  B MMcL also thanks Queen's University Belfast for the award of a Visiting Research Fellowship (VRF).  
PCS and BMMcL thank the US National Science Foundation under the visitors program through a grant to ITAMP
at the Harvard-Smithsonian Center for Astrophysics.   Grants of computational time at the National Energy 
Research Scientific Computing Center in Oakland, CA, USA and at the High Performance Computing Center Stuttgart
(HLRS) of the University of Stuttgart, Stuttgart, Germany are gratefully acknowledged.
%
%
%
%
%
\section*{References}

\providecommand{\newblock}{}

%
%
%
%
%
%

%

\clearpage
%

\clearpage
%

\clearpage
%

%

\clearpage
%

%

%
\clearpage

%

\end{document}